\documentstyle[12pt]{article}
\begin{document}

\hfill{YITPSB-00-77} 

\vspace{1cm}

\centerline{\Large \bf Algebra versus analysis in statistical}
\centerline{\Large \bf mechanics and quantum
field theory}

\vspace{.5cm}

\centerline{\large Barry M. McCoy\footnote{e-mail
mccoy@insti.physics.sunysb.edu}
}
\centerline{\large Institute for Theoretical Physics}
\centerline{\large State University of New York}
\centerline{\large Stony Brook, N.Y. 11794}

\vspace{.5cm}

\begin{abstract}

I contrast the profound differences in  the ways in which
algebra and analysis are used in physics. In particular I discuss the
fascinating phenomenon  that theoretical physicists devote almost 
all their efforts to algebraic problems even though all problems of
experimental interest require some methods of analysis. 

\end{abstract}

\section{Introduction}

When I first started research in physics I was certain that the mathematics used by physics was
analysis. After all, physics was all phrased in the language of
partial differential equations and integrals. We spoke of Fourier
series and
complex analysis and we all learned about norms in Hilbert spaces.
Algebra, if it was considered at all, was concerned with the
classification of finite groups and the determination of which
polynomial equations could be solved in terms of radicals. It was
considered to be a subject of great intellectual interest but not much
practical importance.

But when I reflect on the  many decades which have passed since 
those days of innocence it becomes clear that I have
never genuinely used analysis at all. I have never really cared if
the norm in some space was $L^2$ or $L^1$ and I have certainly never
used the fact that there exist functions which are everywhere
continuous but nowhere
differentiable. In the end all the integrals I ever did were algebraic
and all the special functions I ever used turned out to have  group
theory interpretations. 

Moreover in the years since I was a graduate student there has been
an explosion of knowledge about many body problems (in either
statistical mechanics or quantum field theory) which can be
explicitly solved and in the end the ``solvability'' of these problems
depends on the fact that they have an algebraic structure. Not
coincidently this explosion of knowledge about solvable statistical
mechanical
problems is totally connected with the explosive growth in our
knowledge of algebra in the last 35 years.

The impact of algebra in
physics has been so complete that it now can be said that modern
physics has almost completely abandoned analysis and is now totally
dominated by algebra.

On the other hand there has very recently been a new development in
statistical mechanics which does not seem to fit well into this
algebraic framework. This development is the discovery
by Orrick, Nickel,  Guttmann, and Perk
\cite{n1}-\cite{ongp} that the
magnetic susceptibility of the two dimensional Ising model has a
natural boundary in the complex temperature plane.
I believe that this phenomenon of the natural boundary
is not related to algebra but instead
owes its existence to analysis.

However, the boundary between algebra and analysis is
nebulous and vague and the purpose of this lecture is to clarify the
concepts in a way which makes contact with the way in which
physicists actually treat the subjects. This will be done by
contrasting the following 8 properties and then using these properties
to discuss the new phenomena found for the Ising susceptibility  
and, more generally, how the addition of a magnetic field to the two
dimensional Ising model may turn an algebraic problem into a problem
in analysis.

\bigskip

\begin{tabular}{|l|l|l|}\hline
&{\bf Algebra}&{\bf Analysis}\\ \hline 
1.&Only finite processes allowed&Infinite processes are needed\\
2.& Continuable complex functions& Functions of real variables\\
3.& Integrable systems&Perturbation theory \\
4.&Solvable models&Series expansions\\
5.&S-matrix theory&Field theory\\
6.& Overdetermined holonomic systems&Unstable equations (small denominators)\\
7.& Computable in polynomial time&Not computable in 
polynomial time \\
8.& Simple fixed points &Multiple length  scales\\ 
9.& Ising model at $H=0$&Ising model at $H\neq 0$\\ \hline
\end{tabular}

\bigskip

However it is possible that not all
mathematicians and physicists will be in agreement with
some or all of what I have to say
and to further explain my intentions I will conclude this introduction 
by quoting from the introduction to ``Penguin Island'' written by
Anatole France who discusses corresponding problems in the writing of
history.

{\it ``What is the good, my dear sir, of giving yourself so much
trouble, why compose a history when all you need to do is to copy the
best-known  ones in the usual way? If you have a fresh view or an
original idea, if you present men and things from an unexpected point of
view, you will surprise the reader. And the reader does not like being
surprised. He never looks in a history for anything but the
stupidities that he knows already, If you try to instruct him you only
humiliate him and make him angry. So do not try to enlighten him; he will
only cry out that you insult his beliefs.

``Historians copy from one another. Thus they spare themselves trouble
and avoid the appearance of presumption. Imitate them and do not be
original. An original historian is the object of distrust, contempt,
and loathing from everybody.

``Do you imagine, sir.'' added he, ``that I should be respected and
honored as I am if I had put innovations into my historical works?
And what are innovations? They are impertinences.''}

\section{Finite versus infinite processes}

The first distinction between algebra and analysis which we meet in
our education is that analysis always seems to involve limits and
infinite processes while algebra never involves limits and uses only
finite operations. But even this most elementary distinction can be
misleading and requires at least a passing discussion.

We of course define derivatives as
\begin{equation}
{df(x)\over dx}={\rm lim}_{\Delta \rightarrow 0}{f(x+\Delta
x)-f(x)\over \Delta x}
\end{equation}
and because a limiting process is used we call this analysis. 
But in fact in most elementary courses we actually only consider 
functions such as
\begin{equation}
x^n,~~\sin x,~~e^x,~~  {\rm ln}~ x.
\end{equation}
Thus for example we have
\begin{equation}
{dx^2\over dx}={\rm lim}_{\Delta\rightarrow 0}
{(x+\Delta x)^2-x^2\over \Delta x}={\rm lim}_{\Delta x\rightarrow 0}
{2x\Delta x-(\Delta x)^2\over \Delta x}
={\rm lim}_{\Delta x \rightarrow 0}(2x+\Delta x)=2x
\end{equation}
and even though we have written limit signs in almost every step the
entire computation is nothing but algebra. Indeed from this point of
view most calculus problems are nothing more than an exercise in
algebraic bookkeeping and this is, in fact, how most students treat
their first course in calculus.

We actually do not see much
true analysis until we prove that there exist functions which are
everywhere continuous but nowhere differentiable, 
but this theorem is much too hard
for a beginning course.

We similarly make serious oversimplifications when we first introduce
Riemann integrals as limits of sums. This appears to be a
concept in analysis but again in practice almost  our entire attention
is restricted to integrands which have algebraic or isolated
essential singularities and thus we are usually dealing with a branch
of algebraic geometry without actually admitting it. Furthermore the
 most serious
parts of analysis as used in integration arise only later when 
the concepts of Lebesgue 
measure and Lebesgue integration are introduced.

On the other hand in the last 50 years we have learned that
algebras can be infinite as is seen in following brief history
of the subject

\bigskip 

\begin{tabular}{|l|l|l|}\hline
1870's&Lie&Finite Continuous Groups\\
1944&Onsager \cite{ons}&Infinite loop algebra\\
1968&Kac\cite{kac1}/Moody\cite{moody}&Affine Lie Algebra\\
1985&Drinfeld\cite{drinfeld}/Jimbo\cite{jimbo}&
Quantum finite and affine Lie algebra\\ \hline
\end{tabular} 

\bigskip

\noindent Here, for such objects as affine Lie algebras which have infinite
dimension, the finiteness can be seen for example in the fact that they
can be described by a finite number of generators and relations. In
fact one of the first things done in the theory of these infinite
dimensional algebras is to prove that the definition in terms of the
finite number of generators and relations and the definition in terms
of an infinite number of modes are in fact equivalent \cite{kac}. 
This serves as
a very nice example that something which 
might be thought of as part
of analysis is in fact a part of algebra.

\section{Complex versus real variables}

Closely connected with the distinction between finite and infinite is
the distinction between real and complex variables and here also  
great confusion sets in because of the way we teach calculus. Beginning
students know nothing of complex variables and thus we have no choice
but to consider all variables to be real. But because we only use
algebraic (or trigonometric) functions all of the functions 
we actually use can be
continued to complex variables. The upshot is that we are actually
using analytic functions of a complex variable without ever admitting
it.

This confusion goes much beyond the elementary calculus courses.
Consider the  power series
\begin{equation}
f(z)=\sum_{n=0}^\infty a_nz^n
\end{equation}
which converges in the circle $|z|<R.$ This function $f(z)$ is
analytic inside this circle. In general this is the end of the story
and the function can never be continued outside the circle \cite{hille}.    
The most familiar example of such a function with a natural boundary
is the theta constant
\begin{equation}
\theta_3(0;q)=1+2\sum_{n=1}^{\infty}q^{n^2}
\end{equation}
which has a natural boundary at $|q|=1.$
More generally we may consider series of the form
\begin{equation}
\sum_{k}a_{n_k}z^{n_k}~~{\rm where}~~{\rm
lim}_{k\rightarrow \infty}~k/n_k=0.
\end{equation}
These series are called ``lacunary series'' and they all have natural
boundaries. Another example is the series \cite{hille}
\begin{equation}
\sum_{n=0}^{\infty} \pm c_n z^n
\end{equation} 
where the signs $\pm $ are chosen as independent random
variables. Among these series there is at least one which is
noncontinuable and if $|c_n|^{1/n}\rightarrow 1$ the continuable power
series form at most a countable subset. Hille remarks (page 88 of vol
2 of ref.\cite{hille}) that ``this
fact lends some substance to the assertion that is commonly made that
a power series ``normally'' is noncontinuable.''

Nevertheless even though generically analytic functions have natural
boundaries
most analytic 
functions used by physicists do not have any natural boundaries at all
and can be analytically continued to multi (possibly infinite) 
sheeted Riemann surfaces. These Riemann surfaces almost by definition
have an algebraic structure and are
often related to the representation theory of groups. Furthermore   the
theta functions  which do have natural boundaries are at the foundation
of algebraic geometry. Thus in practice physicists' use of complex
analysis is limited to algebra.

Physicists love complex variables and in particular we love functions
which can be analytically continued. Indeed the very notion of solving a
problem is taken to mean that the answer can be expressed as some
special functions in complex variable theory. We have long ago given
away real variables to the mathematicians, applied mathematicians,
computer scientists, and people who forecast the weather.

\section{Classical integrablity versus canonical perturbation theory}

The meaning and import of these two preceding distinctions between
algebra and analysis is made very concrete in classical mechanics in
the distinction between an integrable system and a perturbation
expansion.

\bigskip

{\bf Definition of Classical Integrability (Liouville 1836)}

A classical system with $2N$ degrees of freedom and a Hamiltonian $H$
is said to be
integrable it there are $N$ independent operators $J_i$ ($i=1,\cdots
,N$) with $J_1=H$ such that 
\begin{equation}
\{J_i,J_k\}=0
\label{cint}
\end{equation}
where $\{A,B\}$ denotes the Poisson Bracket. ( Here when $i$ or $k=1$
 (\ref{cint}) means that $J_k$ is a constant of the motion. The
 remaining equations in (\ref{cint}) say that these constants are all
 compatible (in involution) with each other.)  

\bigskip

These systems may all be solved in closed form by canonical
transformations and have four important properties

\begin{enumerate}

\item These systems have no chaos

\item The dependence on the initial conditions is smooth

\item The equations of the orbits may be continued into the complex plane

\item There is sufficient analyticity in $t$ as $t\rightarrow
\infty$ to determine the asymptotic behavior of the orbit from
(slightly) imprecise initial data.

\end{enumerate}

If Liouville's condition fails to hold then properties 1-4 no longer
hold. However if a system is in some sense ``close'' to an integrable
system we often study it by setting up a ``perturbation'' theory in
some parameter $\lambda$ such that at $\lambda=0$ the system is
integrable. These classical perturbation theories have been studied
for almost 2 centuries and they all have the property that they
never converge because of resonances due to ``small denominators.''
The key to understanding these nonconvergent expansions is the famous
theorem of Kolmogorov, Arnold and Moser (KAM) \cite{kam} which 
dates from the mid
1950's $\rightarrow$ '60's [120 years after Liouville]. This theorem
loosely states that even though commuting constants of the motion do
not exist  the system still has a finite measure of orbits (the KAM
tori) which behave as if they did have all the constants of the
integrable system. However there are also a finite measure of 
chaotic orbits  and these orbits are
infinitesimally close to the KAM tori. When  the ``perturbation''
parameter is made large enough it can happen that these KAM tori
disappear altogether.

\section{Commuting transfer matrices versus series expansions}

In the last 30 years this classical notion of  Liouville Integrability 
has been extended to lattice statistical mechanics and quantum systems 
by means of the
notion of the commuting transfer matrix which was first introduced
into physics with the work of Baxter \cite{bax} on the 8 vertex model. 

A transfer matrix for a statistical system defined on a (say) square
lattice with nearest neighbor interactions builds up the full
partition function, which is defined as
\begin{equation}
Z=\sum_{\rm all~ states}e^{-{\cal E}/kT}
\label{part}
\end{equation}
where $\cal E$ is the interaction energy of the system, by adding one
row of interactions at a time. The matrix T is thus
\begin{equation}
T_{\{\sigma '\},\{\sigma\}}=e^{-{\cal E}(\{\sigma '\},\{\sigma\})/kT}
\end{equation}
where  $\{\sigma ' \}$ and $\{\sigma \}$ denote the configurations of
the variables in 2 adjacent rows and ${\cal E}(\{\sigma
'\},\{\sigma\})$
is that part of the interaction energy which depends only on these two
configurations. This full partition function (\ref{part}) is now given
as
\begin{equation}
Z={\rm Tr} T(u)^{L_h}
\end{equation}
where $L_h$ is the number of rows of the lattice.

As an example we can consider a square lattice where variables
$\sigma=\pm 1$ are on the vertical links and variables $\mu=\pm$ are on
the horizontal links. The transfer matrix which adds one row to the
lattice is then
\begin{equation}
T(\{\sigma
'\},\{\sigma\})=\sum_{\mu_1,\cdots,\mu_N}w(\sigma_1',\sigma_1|\mu_1,\mu_2)
w(\sigma_2',\sigma_2|\mu_2,\mu_3) \cdots w(\sigma_N',\sigma_N|\mu_N,\mu_1).
\label{six}
\end{equation}

\bigskip

{\bf Definition}

A statistical system with a transfer matrix $T$ depending on a
parameter $u$ is said to be 
integrable if 
\begin{equation}
[T(u),T(u')]=0
\label{comm}
\end{equation}

\bigskip

Typically these families of transfer matrices have a value of $u$ 
(which may be taken to be zero for example) such that $T(0)$ is either
the identity or a shift operator. Here we can expand
\begin{equation}
T(u)=T(0)[1+uH+O(u^2)]
\end{equation}
where $H$ will be  of the form
 \begin{equation}
H=\sum_{j=1}^NH_{j,j+1}
\end{equation}
where $H_{j,j+1}$ depends only of the variables on sites $j$ and $j+1$
and $N$ is the number of sites in the row. If we now consider $n^{th}$
derivatives of $T(u)$ with respect to $u$ we see that from (\ref{comm})
we obtain the same condition as the classical Liouville condition (\ref{cint})
except that the Poisson brackets are replaced by commutators.

The condition (\ref{comm}) on the transfer matrix is a global condition
which depends on all the spins in the row. To be fulfilled it is
sufficient for a local condition on the interaction energies to hold.
This local condition is referred to as a star-triangle or Yang-- Baxter
equation. For the case where the transfer matrix is given by 
(\ref{six}) this star triangle equation is
\begin{eqnarray}
& &\sum_{\gamma, \mu'' \nu''} w(\alpha,\gamma|\mu,\mu'')
 w'(\gamma,\beta|\nu,\nu'') w''(\mu'',\nu'|\nu'',\mu')\nonumber\\
 &=& \sum_{\gamma,\mu'', \nu''}
w''(\mu,\nu'|\nu,\mu') w'(\alpha,\gamma|\mu'',\mu') w(\gamma,\beta|\nu'',\nu')
\label{str}
\end{eqnarray}
where the spectral variable in $w,w'$ and $w''$ is $u,u'$ and $u''$
respectively.
Whenever (\ref{comm}) and the corresponding
 star-triangle  holds it has been
possible to exactly compute the eigenvalues of $T(u)$ by algebraic means.
Furthermore for most of these models the order parameters can be
computed. 

The most famous of these models is the Ising model in zero magnetic
field in two dimensions. This model has a variable $\sigma_{i,j}=\pm 1$ at each
site $i,j$ of a lattice with $L_v$ rows and $L_h$ columns which
interact with an interaction energy
\begin{equation}
{\cal E}=-\sum_{i,j}\left( E^v\sigma_{i,j}\sigma_{i+1,j}+
E^h\sigma_{i,j}\sigma_{i,j+1}\right)
\label{isingdef}
\end{equation}
The maximum eigenvalue of the transfer matrix was computed by Onsager
\cite{ons} and from this it is found that in the thermodynamic limit
the free energy per site is (with $N=L_vL_h$) 
\begin{eqnarray}
f&=&-kT \lim_{N\rightarrow\infty} {1\over N}{\rm ln}Z_N\nonumber \\
&=&-kT\left({\rm ln}2+{1\over 2}(2\pi)^{-2}\int_0^{2\pi}d\theta_1\int_0^{2\pi}
d\theta_2{\rm ln }[\cosh 2E_v/kT \cosh 2E_h/kT\right.\nonumber\\
&-&\sinh2E_h/kT\cos\theta_1-\sinh2E_v/kT \cos\theta_2] {\large)}
\label{onsf}
\end{eqnarray}
This free energy has a singularity of the form
\begin{equation}
|T-T_c|^2{\rm ln}|T-T_c|
\label{sing}
\end{equation}
at a temperature $T_c$ defined from
\begin{equation}
\sinh 2E_v/kT_c \sinh 2E_h/kT_c=1.
\end{equation}

On the finite lattice the partition function has no singularities but
instead has zeros in the complex $T$ plane which as 
$N\rightarrow \infty$ lie on the curve
\begin{equation}
\cosh 2E_v/kT \cosh 2E_h/kT
-\sinh2E_h/kT\cos\theta_1-\sinh2E_v/kT \cos\theta_2
\label{zerocurve}
\end{equation}
where $0\leq \theta_1,\theta_2 \leq 2\pi.$
In the limit $N\rightarrow \infty$ this curve of zeroes becomes the
logarithmic singularity and branch cut of (\ref{sing}). 

The spontaneous magnetization (order parameter) of the Ising model is 
\cite{yang2} (for $T<T_c$)
\begin{equation}
M=(1-k^2)^{1/8}
\end{equation}
where
\begin{equation}
k=(\sinh2E_v/kT \sinh 2E_h/kT)^{-1}.
\label{mod}
\end{equation}

It should be remarked that the free energy depends on $E_v/kT$ and
$E_h/kT$ separately while $M$ depends only on the single variable $k.$ 
The variable $u$ of the commuting transfer matrix relation (\ref{comm})
is the variable which parametrises the curve in the plane of $E_v/kT,~
E_h/kT$ on which the $k$ of (\ref{mod}) is constant. This curve may be
expressed in terms of Jacobi elliptic functions where $k$ is the modulus
and $u$ is the variable,

These models with commuting transfer matrices play the role in
statistical mechanics and quantum spin chains which the Liouville
integrable models play in classical mechanics. 
The search for solutions of (\ref{comm}) has
lead to the invention of quantum groups and has been the physics behind
the many discoveries made in algebra in the last 30 years.

On the other hand most problems in statistical mechanics are not of
this integrable form and these models have been studied for over 60
years by means of a variety of series expansions. The first of these
expansions is the expansion of the pressure in terms of the density 
as set up by Mayer \cite{mey} in the 30's. In the 50's the study of
high and low temperatures series expansion was initiated \cite{domb} 
and work on
these expansions continues to the present day. These series expansions
play the role analogous to canonical perturbation theory in classical
mechanics. What is lacking in this analogy of commuting transfer
matrices and series expansions with classical mechanics  
is that we do not know for the
statistical systems what plays the role of the KAM theorem.

\section{S-matrix versus field theory}

In 1967 Yang \cite{yang} studied the S matrix for scattering of
particles in one space and one time dimension 
which interact with a delta function
potential and found that 1) the 2 body S matrix satisfies a set of
overdetermined equations and 2) that all n-body S matrix elements are
determined from the 2 body S matrix. Remarkably enough this
consistency equation for the 2 body S matrix is exactly the
star-triangle equation of Baxter \cite{bax} with the parameter $u$ replaced by
the momentum $p$. It is for this reason that equation (\ref{str}) is
often referred to as the Yang--Baxter equation.

Just as solutions of the star triangle equation may be taken as the
starting point for the study of integrable systems in statistical
mechanics so the Yang--Baxter equation may be taken as the starting
point for the study of solvable S matrix theories in $1+1$ dimensions.
This is the point of view taken by Zamolodchikov \cite{zam} in 1981 in his
initial study of the theory of factorizing S matrices and since then a
large number of these factorizing S matrix theories have been found.
These factorizing S-matrices in $1+1$ dimensions realize the ideas of
the S-matrix bootstrap which originated in work in particle physics
\cite{chew} in the early 60's.
There is obviously a close match (if not a 1-1 correspondence)  between
statistical systems (in 2 dimensions)  with commuting transfer matrices and  
quantum S matrix systems in $1+1$ dimensions.

In contrast with these factorizable S matrices is the concept of
Lagrangian field theory. There are many approaches to this subject but
at least at the computational level the subject is usually treated in
perturbation theory in terms of Feynman diagrams. These Feynman
diagrams plays the role in field theory which canonical perturbation
theory plays in classical mechanics. Just as in classical mechanics   
the perturbation theories do not converge. However, in contrast with
 classical mechanics the analogue of the KAM theorem is not known,

Many (if not most) of the S matrices which satisfy
the Yang Baxter equation may be interpreted as the S matrix for
scattering in some suitable Lagrangian field theory. Thus, for example
it is said that the sine Gordon Field theory has an exactly known S matrix.
On the other hand this identification is made in a curious fashion
because in field theory it is the off mass shell Greens function which
is the object of fundamental interest and with the single exception of
the Ising model these off mass shell greens functions are not known
for any of the models with factorizable S matrices and the agreement
of the S matrix with the Lagrangian is made by arguments such as
saying that they have the same symmetry properties and that they
agree to the one loop level.

There is a folk theorem which says that in theories with factorizable
S-matrices the semi-classical treatment of the corresponding field
theory is exact.

These integrable statistical systems and S matrix theories play the
role in statistical mechanics and field theory that Liouville
integrable systems play    in classical mechanics and their relation
to general systems may be summarized as follows.

\bigskip

\begin{tabular}{|l|l|}\hline
{\bf Integrable}&{\bf Perturbation}\\ \hline
Liouville's Condition&Canonical Orbit Perturbation theory\\
Commuting transfer matrices&Mayer graph expansions\\
& High temperature expansions\\
S-matrix theory & Feynman Diagrams\\ \hline
\end{tabular}

\section{Holonomic and D finite systems}

\bigskip

{\bf Definition of holonomic }

\bigskip 

A system of partial differential equations is said to be holonomic 
\cite{kk} or
maximally overdetermined if only a finite number of initial conditions
(instead of a function) need to be specified  to uniquely determine the
solution. 

\bigskip

This notion of holonomic is the multivariable generalization of the
notion of D-finiteness.

\bigskip 
{\bf Definition of D-finite }

\bigskip

A function $f(z)$ is said to be D(ifferentially) finite \cite{lip} if there exists
an integer $k$ and polynomials $P_0(z),\cdots,P_k(z)$ such that
$P_0(z)\neq 0$ and
\begin{equation}
0=P_0(z)f(z)+P_1(z)f'(z)+\cdots P_k(z)f^{(k)}(z)
\end{equation}

In 1977 Kashiwara and Kawai \cite{kk} proved the remarkable  
theorem that every Feynman diagram is a holonomic function of teh
external momentum.
However, even though this is true for each diagram in the perturbation
expansion the complete sum of all the diagrams is not in general holonomic.
 
On the other hand there are field theories where the entire Greens
function is still holonomic. This is illustrated by the field theory
based on the Ising model and is the reason that the series of papers
of Sato, Miwa, and Jimbo \cite{smj} on the Ising model is entitled ``Holonomic
Quantum Field Theory.'' It is strongly suspected that all of the field
theories which have factorizable S-matrices are holonomic field
theories.
These field theories then are clearly analogous to the integrable
systems of classical mechanics and the fact that the full Greens
function does not share the property of being holonomic with the
individual Feynman diagrams seems     to be related to the fact that
in classical mechanics the
canonical perturbation theory does not converge and that 
a finite fraction of the orbits will not lie on $N$ dimensional KAM tori.

Furthermore there is an important connection between the concepts of
D-finiteness  and the absence of natural boundaries. In particular let
\begin{equation}
 F(x,y)=\sum_{n\geq 0} y^n H_n(x)
\end{equation}
be a D-finite series in $y$ with rational coefficients. For $n\geq 0$
let $S_n$ be the set of poles of $H_n(x)$ and let $S=\cup _nS_n.$ Then
the set $S$ has only a finite number of accumulation points and thus
has no natural boundaries \cite{theorem}. This strongly suggests that natural
boundaries (if they arise at all) will not be present in any finite
order or perturbation theory but will only occur in the full sum.

\section{Polynomial versus NP complete systems}

Thus far I have contrasted algebra and analysis by using concepts
familiar to most physicists and mathematicians. But an even broader
comparison of the two subjects can be made if we extend our point of
view to embrace computer science. Here an important concept  
is NP completeness.

The determination of the  time it takes to run an algorithm on a computer is
obviously of great importance. In particular if we have a problem
(such as the computation of a partition function or free energy per 
site) which depends on
a number $N$ of ``input parameters''  
it is of great practical importance how the running time of the
algorithm increases with $N$. If the running time increases as a power
of $N$ the problem is said to be solvable in polynomial time.
For a lattice statistical mechanical system such as the Ising model
(\ref{isingdef}) the number $N$ is the number of lattice points in the system.

Unhappily there are very few problems which physicists are interested
in which have been shown to be solvable in polynomial time. What
almost always happens is that the running time increases exponentially
(or worse) as the size of the system increases. This unfortunate fact
is known to every physicist whoever tried to do a numerical computation
on a large system.

   Computer scientists deal all the time with difficult problems for which
   no polynomial-time algorithm has been found, despite decades of attempts;
   being immodest, they strongly (and probably rightly) suspect that
   no polynomial-time algorithm for these problems exists.  A famous example
   is the Traveling Salesman Problem:  given $N$ cities on the plane,
   determine whether there exists a tour of total length $\le L$.
   The naive approach requires checking $N!$ possible tours;  and while
   better algorithms have been found, the $N!$ has never been reduced to $N^p$.
   Note, however, that if someone claims to {\em exhibit}\/ a tour of
   length $\le L$, this claim can trivially be checked in polynomial
   (in fact linear) time.  A problem for which a purported solution is
   {\em checkable}\/ in polynomial time is said to belong to the family NP.
   (The letters stand for Nondeterministic Polynomial.  The idea is that
   if a purported solution is polynomial-time checkable, then the original
   problem is solvable in polynomial time by a ``{\em nondeterministic}\/
   Turing machine'' that could test all possible solutions {\em in parallel}\/.)
   Now, among the problems lying in NP, computer scientists distinguish a
   subclass of problems called {\em NP-complete}\/:  these are the most
   difficult problems in the class NP, in the sense that if one NP-complete
   problem were solvable in polynomial time, then {\em all}\/ problems in NP
   would be.  And literally hundreds of problems --- the Traveling Salesman,
   graph $q$-colorability, \ldots --- have been proven rigorously to be
   NP-complete.  So, either all of these problems are polynomial-time solvable
   (P=NP), or none of them are (P$\neq$NP).  Computer scientists strongly
   suspect that the latter is true, and their Holy Grail is to prove it
   \cite{garey,clay}.

On the other hand when a physicist refers to ``solving a problem''
he/she almost always means that the problem has been reduced to an
algebraic problem in the sense I have discussed above. This algebraic
structure almost certainly will allow the problem to be solved in
polynomial time. 

There is obviously a gap between the computer science concepts of 
polynomial time versus NP complete and the physicist's concept of
integrable versus nonintegrable. In particular it is not clear that
there is a one-to-one match between the term NP complete and the term
nonintegrable. Nevertheless some physics problems 
have become so well known
that they have attracted the attention of computer scientists and
in the past 15 years Ising spin glasses \cite{bar} 
and the monomer dimer problem \cite{jer} have been shown to be NP complete. 
The most recent of these studies is by Istrail \cite{ist} in a
paper (accompanied by a press release) entitled {\it Statistical
Mechanics, Three Dimensionality and NP-Completeness 1. Universality
of intractability for the partition function of the Ising model across
non-planar lattices}. This paper suggests (but to the best of my
reading does not actually say) that 

\bigskip

{\bf (Suggestion) The free energy per site of the three dimensional
Ising model  on a homogeneous lattice is NP complete.}

\bigskip 
\noindent Istrail interprets his suggestion to imply that it is
impossible to ``solve'' the three dimensional Ising model.

This may or may not be considered surprising depending on your point
of view. For over 50 years people have repeatedly tried and failed to find a
solution to the three dimensional Ising model which more or less
generalizes Onsager's result (\ref{onsf}). In addition people have
searched for and failed to find families of commuting transfer matrices which
include the three dimensional Ising model. All of this has led people
to suppose that a ``solution'' does not exist. The suggestion of Istrail fits
nicely into this body of experience.

But what one would like is a more constructive discussion of the
problem. For example one would like to know if by
looking at the high temperature series expansion of the free energy
there is any way to see some difference between the 2 and 3
dimensional model. The most standard high temperature series expansion
is to consider the isotropic case with only one interaction energy
$E=(E_v=E_h~{\rm in~two~ dimensions})$ and use $v=\tanh E/kT$ as the
expansion variable. Then in dimension $d$ the (exponential of the)
free energy can be written in
terms of $v$ as
\begin{equation}
e^{-f/kT}= 2\cosh^d E/kT\left(1+\sum_{n=2}^{\infty}a_{2n}v^{2n}\right).
\label{hts}
\end{equation}

In two dimensions the coefficients $a_n$ may be quickly obtained from
the exact solution (\ref{onsf}) to as large an $n$ as
desired. For three dimensions the best results available \cite{ge}
only go up to $N=26$

\begin{tabular}{|r|r|} \hline \\ 
order n& the coefficients $a_n$\\ \hline
4&3\\ 
6&22\\
8&192\\
10&2,046\\
12&24,853\\
14&329,334\\
16&4,649,601\\
18&68,884,356\\
20&1,059,830,112\\
22&16,809,862,992\\
24&273,374,177,222\\
26&4,539,862,959,852\\ \hline
\end{tabular}

\bigskip

A finite number of terms from a power series expansion will only define
a polynomial and thus in strict principle we can never learn anything
about singularities in the free energy from the 12 coefficients given
above without further assumptions. But as physicists we ``do the best we
can'' and  accordingly   
high temperature series expansions for free energies (and specific
heats) are inevitably analyzed by assuming that there is a critical
value of $v_c$ (at the radius of convergence of the full infinite
 series (\ref{hts})
such that at $v_c$ there is a singularity of the form
\begin{equation}
f\sim A(v_c-v)^{2-\alpha}~~{\rm or}~A(v_c-v)^2{\rm ln}|v_c-v|
\label{assing}
\end{equation}
which generalizes (\ref{sing}). But even for the two dimensional case
the expression (\ref{assing}) is not the exact answer  so several further 
terms which are analytic at $v_c$ 
and possibly a few further confluent singularities at $v_c$ 
are assumed to exist and the
resulting form is fitted to the series expansion by some technique
such as differential Pad{\'e} approximants. 

Ultimately this procedure of extracting a critical exponent from the
high temperature series data  depends on  the form of the singularities
we have assumed to fit the data and can be thought of as a
nondeterministic step. But in section 4 we pointed out that
most power series do not lead to such simple algebraic (or
logarithmic) functions and
that in general natural boundaries are to be expected. Thus 
generically the assumptions made to analyze a finite number of terms
in high temperature series
expansions can fail to be correct for the full infinite series. 
To the extent that it is felt that the three dimensional Ising model
is ``computationally intractable'' and NP complete so it would seem to
become increasingly unlikely that the specific heat should be described
by such a very special and nongeneric form as (\ref{assing})

I hasten to point out that there is no known  useful \
neccesary and sufficient condition for the
coefficients of a power series expansion which tell  if a natural
boundary is present. Moreover it is not expected that 
the first 12  terms would even come close to approximating this criterion
even if a criterion were known. 
Nevertheless there is a genuine problem here which demands to be
studied. 

\section{Simple fixed points}

The free energy, spontaneous magnetization and correlation functions
of the two dimensional Ising model at $H=0$ are all exactly known and
the model has a phase transition at the critical temperature $T_c.$
Moreover the properties at this phase transition are all in qualitative
agreement with the properties of real phase transitions as seen in
experiments even though for these real systems we are unable to
compute the phase transition behavior theoretically. Therefore
starting in the mid '60's a very successful phenomenology of phase
transitions was developed and converted into a physical theory in the
70's which says in effect the the real world
behaves like the two dimensioal Ising model at $H=0$. 

The crucial ingredient in this description is the notion of a single unstable
fixed point which plays a key role in the theory of the
renormalization group \cite{kad}-\cite{fis}.

The correlation functions of the 2-dimensional Ising model at $H=0$ have
the property that if $T\neq T_c$ the correlations (on a lattice of
unit spacing) decay at large separations as
\begin{equation}
<\sigma_{0,0}\sigma_{\vec R}>\sim M^2+A(\theta)|R|^{-p}e^{-|R|/\xi(T,\theta)}
\end{equation}
where $\theta$ is the angle which ${\vec R}$ makes with the $x$ axis and
$p=1/2$ if $T>T_c$ and 2 is $T<T_c.$ If the system is isotropic
$(E_v=E_h)$ and if $T\rightarrow T_c$ the dependence on $\theta$ disappears
and so we will suppress it in our further discussion.
The quantity $\xi(T)$ is called the correlation length and it has the
property that as $T\rightarrow T_c$
\begin{equation}
\xi(T)\sim |T-T_c|^{-\nu}~~{\rm with}~\nu=1
\end{equation}

At $T_c$ the two point function of the Ising model behaves for large
$|R|$ as
\begin{equation}
<\sigma_{0,0}\sigma_{\vec R}>_{T=T_c}\sim{C\over |R|^{d-2+\eta}}~~{\rm
with}~ \eta=1/4
\label{siglarg}
\end{equation}

\bigskip 

{\bf Definition of scaling limit}

The scaling limit is the limit in which
\begin{eqnarray}
& &|R|\rightarrow \infty,~~T\rightarrow T_c\\
& &{\rm with}~ |R|/\xi(T)=|R||T-T_c|^{\nu}=r~~{\rm fixed}
\end{eqnarray}

\bigskip

{\bf Definition of Scaled two point function}

The scaled two point function (for $T$ either above or below
$T_c$) is defined as
\begin{equation}
G_{\pm}(r)={\rm lim} M_{\pm}^{-2}<\sigma_{0,0}\sigma_{\vec R}>
\label{scaling}
\end{equation}
where 
\begin{equation}
M_{\pm}=(1-k^{\mp2})^{1/8}
\end{equation}
(and we note that $M_{-}$ is the spontaneous magnetization).
These scaled two point functions \cite{wmtb} exist and are nonzero for the 2
dimensional Ising model at $H=0.$

\bigskip

{\bf Definition of simple fixed point scaling}

If the behavior of $G_{\pm}(r)$ as $r\rightarrow 0$ is
\begin{equation}
G_{\pm}(r)\sim {C'\over r^{d-2+\eta}}
\label{gsmall}
\end{equation}
where (1) the $\eta$ of (\ref{siglarg}) and (\ref{gsmall}) are the same
and (2) the constant $C'$ is obtained from $C$ by using the definition
(\ref{scaling}) then the system is said to have  simple fixed point scaling. 

\bigskip

These two defining properties of simple fixed point  scaling have been
verified for the two dimensional Ising model. The equality of the
exponents $\eta$ was demonstrated \cite{mtw} in 1977 and the equality
of the constants was demonstrated \cite{cat} in 1991. But these
computations rely completely on all of the special properties of the
Ising model which lead to the Painlev{\'e} representation of the scaled
two point function \cite{wmtb}. There are no other integrable models
for which the corresponding computations have been yet carried
out. Therefore it is totally correct to say that the two dimensional
Ising model in zero magnetic field is the only system for which simple
fixed point scaling has ever been proven to hold.

It is, however, usually assumed that this simple fixed point 
scaling will hold
for all integrable models coming from a family of commuting transfer matrices.
Moreover it is assumed much more generally that this scaling holds in
the general theory of critical phenomena \cite{kad}--\cite{fis}
and in particular it is
assumed to hold for the three dimensional Ising model and for real fluids 
such as noble gases and ${\rm CO}_2.$

Indeed this assumption of single fixed point scaling is much older than
the theory of critical phenomena. It is used in the  renormalization
theory in quantum field theory in the work of Gell--Mann and Low
\cite{gell} on short distance behavior in QED and was later carried
forward by Callan
\cite{call} and Symanzik \cite{sym} in what is now called the
Callen-Symanzik equation. If this simple fixed point assumption fails
then many of the results which are obtained from
the renormalization group will break down.

Because of this very widespread use of results which follow from
single fixed point scaling it is of 
great importance find an appropriate
justification for it. Thus it is somewhat disturbing that for the
Ising model this scaling has only been proven in ref. \cite{mtw} and
\cite{cat}  by using very strong
integrability properties. These integrability properties are certainly
algebraic in the sense I have discussed above and therefore it is
believable that eventually we will be able to prove analogous
results for other integrable  statistical mechanical  models. 

But if the existence of single fixed point scaling in integrable
models does indeed require the use of an algebraic structure then
this very fact casts doubt on whether they can hold for a non
integrable system such as the three dimensional Ising model,
a simple fluid, or an interacting quantum field theory such as QED. 

It would seem quite possible that for non integrable systems the
exponent part of the scaling law could hold while the equality of the
constants could fail. Even worse it could happen that as
$r\rightarrow 0$ the Green's  function could behave as
\begin{eqnarray}
G(r)&\sim& {{\rm ln}^a r\over r^{d-2+\eta}} ~~{\rm or}\nonumber\\
 &\sim&{\rm ln}^a r 
{\rm ln}~{\rm ln}^b r/r^{d-2+\eta}\\
\end{eqnarray}
where each logarithm effectively introduces another scale. Since such
logarithmic ``shock wave'' layers are known to happen \cite{as} in the Korteweg de
Vries equation which is an integrable partial differential equation it
would seem perfectly possible that multiple  scales
could exist in a non integrable system. 

We are thus lead to the following

\bigskip

{\bf Major Question}

\bigskip

{\bf Does single fixed point scaling hold for all systems or does it only
hold for the integrable (algebraic) systems?}

\section{The Ising model for $H\neq 0$}

Now that I have explored in detail the many different ways in which
analysis and algebra are used in physics  I may
return to the topic which originated this discussion in the first
place;
The study of the susceptibility of the two dimensional Ising model at
$H=0$ made in references \cite{n1}-\cite{ongp}.

The magnetic susceptibility is the derivative of the magnetization and
is expressed in terms of the two point correlation function as
follows:
\begin{equation}
kT\chi=kT{\partial M(H)\over \partial H}|_{H=0}
=\sum_{j,k}\left(<\sigma_{0,0}\sigma_{j,k}>-M^2\right)
\end{equation}
where by the interaction of the Ising model with the magnetic field $H$
we mean the addition of the term $-H\sum_{j,k}\sigma_{j,k}$ to the
interaction energy (\ref{isingdef}). 

The susceptibility is different for $T$ above and below $T_c$ and in
ref.\cite{wmtb} it is shown that
\begin{eqnarray}
kT\chi_{+}(T)&=&k(1-k^{-2})^{1/4}\sum_{l=0}^{\infty}{\hat
\chi}^{(2l+1)}\nonumber\\
kT\chi_{-}(T)&=&(1-k^2)^{1/4}\sum_{l=1}^{\infty}{\hat \chi}^{(2l)}
\label{sus}
\end{eqnarray}
where ${\hat \chi}^{(n)}$ is the sum over $(j,k)$ of the $n$ particle
contribution to the two point function.

For finite $n$ each term ${\hat \chi}^{(n)}$ in (\ref{sus}) is a 
holonomic function of
$T$. But in ref. \cite{n1} and \cite{ongp} compelling evidence
(just short of a proof) is given which indicates that the full sum
$\chi_{\pm}(T)$ is not a holonomic function but has a natural boundary
on the same curve (\ref{zerocurve}) on which the partition function on
the finite lattice has zeros.
This natural boundary is a new phenomenon.

But once it is accepted that the susceptibility at $H=0$ has a natural
boundary then it is to be expected that for all $H\neq 0$ the two
dimensional Ising model can have a natural boundary on the curve where
the partition function on the finite lattice has zeroes. 
This is in agreement with the intuition suggested by ref.\cite{jer},
which proved that the two--dimensional monomer dimer problem is NP
complete, because the Ising model in a magnetic field can be reduced
to this monomer dimer problem.

We thus see that there are sound reasons for believing that the Ising
model in a magnetic field is a problem in analysis and not of
algebra. Therefore I pose the question:

\bigskip

{\bf Does simple fixed point scaling fail for the two dimensional Ising
model with $H\neq 0$?}

\bigskip

The situation, however, is more subtle that this preceding
discussion indicates because in 1989 Zamolodchikov \cite{zam2} showed that in
the scaling limit there is a continuum S-matrix model  in the same
universality class as the Ising model at $T=T_c,~H>0$ which is
integrable and this S matrix model has subsequently been shown to
arise from an integrable lattice model \cite{wns} (which however is 
not the lattice
Ising model at $T=T_c~H>0.$) The relation between these two models, one
of which is integrable and the other which is not, is not understood. 

\section{Conclusions}

I have now completed the survey of eight properties which can be used
to characterize the difference between algebra and analysis in
statistical mechanics (and quantum field theory) and have 
given one concrete example where at least some of 
these differences can be seen to occur. Several of the questions raised
about the possible breakdown of scaling contradict beliefs long held
by many researchers in both statistical mechanics and quantum field
theory and may be classified as impertinences in the sense
contemplated by Anatole France. 
But impertinent as some of my distinctions between algebra and
analysis are, there are no known theorems which prove them to be incorrect. 

Moreover when taken together the survey presented here is
very disturbing because almost all our theoretical computations and
intuition come from algebraic problems whereas all real problems of
experimental interest will not possess these algebraic structures.  
This leads to the nagging suspicion that there must be phenomena which
are left out of our phenomenological picture of critical phenomena and
quantum field theory. 

One property of nature which manifestly is left out of all solvable
problems in statistical mechanics and quantum field theory is the
phenomenon  of particle production and the converse phenomenon of decay
of an unstable particle.
This is very relevant to Zamolodchikov's model \cite{zam}
because it has a mass spectrum of 8 particles of which 5 are above the
2 particle decay threshold. Therefore when the model is perturbed away
from $T=T_c$ these 5 particles will decay (at least using perturbation
theory \cite{mus}).

But what is meant by an unstable particle in a mathematical sense? One
common definition is to continue the 2 point function as a function of
momentum through the 2 particle cut onto what is called the second
sheet and to identify an unstable particle as a pole on this second
sheet. But in deriving the analytic structure of the two point
function for the Ising model we make an expansion  
in terms of the complete set of eigenvectors and eigenvalues of the
transfer matrix. For the Ising model these eigenvalues are very smooth
for a finite size system and they go over smoothly to the usual
kinematic cuts in the thermodynamic limit. But once the system becomes
non integrable it is known from computer studies of the eigenvalues
that what looks like randomness creeps into the eigenvalues due to
``avoided crossings.'' For the same reason that ``noisy''
coefficients in a power series expansion could generate natural
boundaries it would seem  that as a function of
$k$ the two point function could also have a natural boundary and
cannot be continued
onto a second sheet. If this is the  case the definition of unstable particle
as a pole on the second sheet cannot be used.

What is needed is some way to quantify the influence of a nonintegrable
perturbation on an integrable statistical mechanical system. In other
words at the very least 

\bigskip

{\bf We need a statistical mechanical analogue of the KAM theorem.}

\bigskip

One possible version of such a theorem would be to prove that in the scaling
limit the intuitive ideas of the renormalization group
\cite{wk}-\cite{fis} are exact. For example it would seem not
unreasonable to expect that the addition of bonds to the Ising model
which destroy the planarity of the lattice (such as next nearest
neighbor interactions) will have absolutely no effect on the scaling
function even though the arguments of Istrail \cite{ist} would lead one to
believe that the system is NP complete. 

The fact is that 25 years
after the formulation of the renormalization group no such rigorous
theorem has been proven even though such a theorem is sorely
needed if we are to understand how algebra and analysis are related in
the study of physical systems. In the end it is ultimately unacceptable
to ignore the fact that algebraic systems are only a set of measure
zero in the space of all systems. No matter how much we physicists love
algebra we cannot ignore the fact that analysis exists.

\vspace{.3in}

{\Large \bf Acknowledgments}

\bigskip

I am very pleased to acknowledge many useful discussions with
A. Guttmann, W. Orrick and A. Sokal.
This work is supported in part by NSF grant DMR0073058.

\end{document}